\documentclass[aps,prc,groupedaddress]{revtex4}
\usepackage{graphicx,bm}
\usepackage{amsmath,bm}

\DeclareMathOperator{\nrd}{\overleftrightarrow{\nabla}}
\DeclareMathOperator{\nr}{\overrightarrow{\nabla}}
\DeclareMathOperator{\vecsigma}{\overrightarrow{\sigma}}

\begin{document}

\title{Subleading contributions to the three-nucleon contact interaction}
\author{L.\ Girlanda$^{\,{\rm a}}$, A.\ Kievsky$^{\,{\rm b}}$,
and M.\ Viviani$^{\,{\rm b}}$}
\affiliation{
$^{\,{\rm a}}$\mbox{Dipartimento di Fisica, Universit\`a del Salento, and INFN Sezione di Lecce,}\\ \mbox{Via Arnesano, I-73100 Lecce, Italy}\\
$^{\,{\rm b}}$\mbox{INFN, Sezione di Pisa, Largo Bruno Pontecorvo 3, I-56127 Pisa, Italy}\\
}

\date{May 6, 2020}

\begin{abstract}
We obtain a minimal form of the two-derivative three-nucleon contact Lagrangian, by imposing all constraints deriving from discrete symmetries, Fierz identities and Poincar\'e covariance. The resulting interaction, depending on 13 unknown low-energy constants, leads to a three-nucleon potential which we give in a local form in configuration space. We also consider the leading (no-derivative) four-nucleon interaction and show that there exists only one independent operator.
\end{abstract}


\maketitle

\section{Introduction}
Purely contact interactions  are crucial components of the
nucleon-nucleon ($NN$) and multi-nucleon forces as derived in chiral
effective field theory ($\chi$EFT) \cite{bernard95,vankolck99,bedaque02,epelbaum06,epelbaum09}. They encode the short distance properties of the nuclear interaction as opposed to terms involving pion exchanges, which are of larger range. At sufficiently low energy, even the pion can be integrated out, giving rise to the pionless effective theory.  The contact vertices are the same in both versions of the theory, the only difference being in the value of the accompanying low-energy constants: in the pionless theory the latter implicitly include the effect of pions, considered as heavy particles.  At the level of the $NN$ system two
purely contact terms appear at leading order (LO) and contribute to
the central part of the interaction \cite{weinberg90,weinberg91,ordonez92,ordonez94,ordonez96}. In the three-nucleon ($3N$)
system  the first non-vanishing contribution appears at
next-to-next-to-leading order (N2LO) in the chiral expansion \cite{vankolck94,epelbaum02}.  
 This term has already been included in some three-nucleon interactions
(TNI) as the $E$-term \cite{navratil,gazit}. Indeed the inclusion of such term was 
found to be mandated, in the framework of pionless EFT, by the requirement of
 renormalizability \cite{Bedaque:1998kg, Bedaque:1998km, Bedaque:1999ve, 
Hammer:2000nf}. In the present paper we focus on the subleading $3N$ contact terms, i.e. those containing two space-derivatives of nucleon fields, and stick to isospin symmetric operators. 
Only one such term was found to be necessary in Ref.~\cite{Bedaque:2002yg} in order to achieve cutoff independence (see however Refs.~\cite{Platter:2006ev, Platter:2006ad} for a different claim). We would like to stress however, that in the EFT framework one has to include all terms allowed by symmetry at a given order, not just the one needed by the renormalization procedure, since the EFT is the most general theory encoding the given symmetry properties. 
The plan of the paper is the following: in Section \ref{sec:lo} we illustrate our strategy to determine a minimal set of $3N$ contact operators by considering the leading non-derivative operators, thus reobtaining the result that only one operator arises at this order; in Section \ref{sec:nlo} the same strategy is applied to the list of all possible two-derivative operators, thus reducing their number from 146 down to 18; further constraints from relativity, as pertinent to momentum-dependent interactions, are discussed in Section~\ref{sec:rel}, and allow to reduce the number of independent operators to 13; we give in Section~\ref{sec:pot}  the resulting $3N$ potential in coordinate space in a local form, by choosing an appropriate momentum cutoff. Finally, in Section~\ref{sec:4n} we consider the leading four-nucleon ($4N$) contact Lagrangian and find that there exists only one such operator.

\section{Fierz constraints on the leading contact interaction} \label{sec:lo}
Rotational, isospin and time-reversal invariance allows to list 6
possible operators describing three-nucleon contact 
interactions without derivatives.
Indeed, since $N^\dagger {\bm \sigma} N$ is odd under time-reversal,
an even (odd) number of $\sigma$ matrices has to be associated with a
purely real (imaginary) isospin structure. 
Therefore, the leading order\footnote{The word "leading", in this context, refers to all possible $3N$ contact Lagrangians. As already mentioned, the resulting interaction contributes at the N2LO in the chiral expansion.}
 $3N$ contact  Lagrangian density reads 
\cite{epelbaum02}:
\begin{eqnarray} \label{eq:lolagr}
{\cal L}_{\mathrm{3N}}^0&=&-E_1 N^\dagger N N^\dagger N N^\dagger N
- E_2 N^\dagger \sigma^iN N^\dagger \sigma^i N N^\dagger N
\nonumber \\
&&- E_3 N^\dagger \tau^a N N^\dagger \tau^a N N^\dagger N 
- E_4 N^\dagger \sigma^i\tau^a N N^\dagger \sigma^i \tau^a N
N^\dagger N 
\nonumber \\
&&- E_5 N^\dagger \sigma^i N N^\dagger \sigma^i \tau^a N N^\dagger 
\tau^a N
- E_6 \epsilon^{ijk} \epsilon^{abc} N^\dagger \sigma^i \tau^a N
N^\dagger \sigma^j \tau^b N N^\dagger \sigma^k 
\tau^c N \nonumber \\
&\equiv&  -\sum_i^6 E_i O_i^{(0)}
\end{eqnarray}
Symmetry properties under permutations are encoded in the
anticommuting nature of the nucleon field $N$. The properties under
exchange of spin  indeces can be conveniently expressed as
\begin{equation}
\begin{array}{l}
({\bf 1} ) [ {\bf 1} ] = \frac{1}{2} ({\bf 1} ] [ {\bf 1} ) +
      \frac{1}{2} ({ {\bm \sigma}} ] \cdot [ { {\bm \sigma}} )\\
( \sigma^i ) [{\bf 1}] = \frac{1}{2} ( \sigma^i ] [ {\bf 1}) +
      \frac{1}{2} ({\bf 1} ] [ \sigma^i ) - \frac{i}{2} \epsilon^{ijk}
      ( \sigma^j ] [ \sigma^k)\\
(\sigma^i ) [\sigma^j] =\frac{1}{2}\left\{\delta^{ij} ({\bf
        1}][{\bf 1})-\delta^{ij} ({\bm \sigma}] \cdot [{\bm \sigma}) +
      (\sigma^i ] [\sigma^j) + (\sigma^j] [\sigma^i)+ i \epsilon^{ijk}
      ( \sigma^k ] [ {\bf 1} ) - i \epsilon^{ijk} ( {\bf 1 } ] [
      \sigma^k ) \right\},
\end{array} 
\end{equation}
where ${\bf 1}$ is the identity operator in the one-particle spin space, and (,) and [,] denote spin  indeces of the enclosed
operator. Similar relations hold in the one-particle isospin space for the identity and $\tau^i$ operators. Simultaneous Fierz rearrangements of spin and isospin
indeces of nucleon fields 1--2, 1--3 and 2--3 allow then to derive a
set of  linear relations among the above operators, in a similar way
as was done for the (parity violating) two-nucleon contact interaction in Ref.~\cite{pvlagr}.
For instance, from the permutation of nucleons 2--3 we find,
\begin{equation}
\begin{array}{l}
O_1^{(0)} = -\frac{1}{4} \left( O_1^{(0)} + O_2^{(0)} + O_3^{(0)} +
O_4^{(0)} \right) \\
O_2^{(0)} = -\frac{1}{2} \left( O_2^{(0)} + O_5^{(0)} \right) \\
O_3^{(0)} = -\frac{1}{2} \left( O_3^{(0)} + O_5^{(0)} \right) \\
O_4^{(0)} = -\frac{1}{4} \left( 2 O_4^{(0)} + 2 O_5 ^{(0)} - O_6^{(0)}  \right) \\
O_5^{(0)} = -\frac{1}{2} \left( 3 O_2^{(0)} - O_5^{(0)} \right) \\
O_6^{(0)} = 2 \left(  O_4^{(0)} - O_5^{(0)} \right),
\end{array}
\end{equation}
leaving only one independent operator (in agreement with
Ref.~\cite{epelbaum02}). The
relations deriving from the remaining permutations are linearly
dependent on the ones above.
The remaining operator was chosen to be $O_3^{(0)}$ in Ref.~\cite{epelbaum02}, and it gives rise, in momentum space to a contact potential 
\begin{equation}
V_{\mathrm{cont}}=\frac{1}{2} \sum_{j\neq k} E {\bm \tau}_j\cdot {\bm \tau}_k,
\end{equation}
with $E$ denoting the corresponding low-energy constant (LEC). Choosing a cutoff depending on momentum transfer, it is possible to obtain the corresponding coordinate space potential in a local form~\cite{navratil}. A sensitivity study of bound state and scattering observables in $A\leq 4$ systems to this and other components of the TNI was  performed in Ref.~\cite{3nforces}.
It should be noticed that the equivalence in the choice of the contact
operator is true at the 
Lagrangian level, but it is in general spoiled at the level of the 
$3N$ potential by the cut-off, which may involve non-symmetrical
combinations of the nucleon momenta.
At any rate, in the effective theory framework, different choices are
equivalent up to higher order corrections.

\section{Fierz constraints on the subleading contact interaction} \label{sec:nlo}
Parity requires that the next-to-leading order $3N$ contact Lagrangian contain two
spatial derivatives. Using translational invariance (or momentum
conservation) the only possible space-structures can be taken to be of
the form
\begin{equation}
\begin{array}{l}
X^{+}_{A,ij} = (N^\dagger \nrd_i N)  (N^\dagger \nrd_j N) (N^\dagger N)\\

X^{+}_{B,ij} =\nabla_i (N^\dagger N) \nabla_j (N^\dagger N) (N^\dagger N)\\
X^{-}_{C,ij} =i \nabla_i (N^\dagger N)  (N^\dagger \nrd_j N) (N^\dagger N)\\
X^{+}_{D,ij} = (N^\dagger \nrd_i \nrd_j N)  (N^\dagger N) (N^\dagger N),
\end{array}
\end{equation}
where the $i$ is required by the hermiticity condition and (hereafter)
the superscripts denote the time-reversal properties. 
The relevant isospin structures are
\begin{equation}
T^+={\bf 1}, \quad
{\bm \tau}_1  \cdot {\bm \tau}_2, \quad
{\bm \tau}_1  \cdot {\bm \tau}_3, \quad
{\bm \tau}_2  \cdot {\bm \tau}_3, \quad
T^-={\bm \tau}_1 \times {\bm \tau}_2 \cdot {\bm \tau}_3,
\end{equation}
where the subscripts of the Pauli matrices refer to the nucleon
bilinears they belong to.
Even (odd) combinations of $X \otimes T$ structures under time-reversal have to be
associated with spin structures containing even (odd) numbers of
$\sigma$ matrices. Finally, the spin-space indeces have to be
contracted with Kronecker $\delta$'s or Levi-Civita tensors
$\epsilon$'s.
Following the above procedure we have obtained a list of 146
operators, displayed in Table~\ref{tab:operators}.
\begin{table}
\begin{scriptsize}
\begin{tabular}{|l|c||l|c|}
\hline
$o_{1-3}$ & $\nrd_1 \cdot \nrd_2  [ {\bf 1}, {\bm
    \tau}_1\cdot {\bm \tau}_2, {\bm \tau}_1\cdot {\bm \tau}_3]$
&
$o_{73}$ & $i \nrd_1 \cdot \vecsigma_3 \nr_2 \cdot \vecsigma_2
 [{\bm \tau}_1 \times {\bm \tau}_2 \cdot {\bm \tau}_3]$ 
\\
$o_{4-6}$  & $\nrd_1 \cdot \vecsigma_1 \nrd_2 \cdot
\vecsigma_2  [ {\bf 1}, {\bm
    \tau}_1\cdot {\bm \tau}_2, {\bm \tau}_1\cdot {\bm \tau}_3]$
&
$o_{74}$ & $i \nrd_1 \cdot \nr_2 \vecsigma_2\cdot \vecsigma_3
[ {\bm \tau}_1 \times {\bm \tau}_2 \cdot {\bm \tau}_3]$ 
\\
$o_{7-9}$  & $\nrd_1 \cdot \vecsigma_2 \nrd_2 \cdot
\vecsigma_1  [ {\bf 1}, {\bm
    \tau}_1\cdot {\bm \tau}_2, {\bm \tau}_1\cdot {\bm \tau}_3]$
&
$o_{75-78}$ & $i \nrd_1\times\nr_2\cdot\vecsigma_1 
  [ {\bf 1}, {\bm
    \tau}_1\cdot {\bm \tau}_2, {\bm \tau}_1\cdot {\bm \tau}_3,{\bm \tau}_2\cdot {\bm \tau}_3]$
\\
$o_{10-12}$  & $\nrd_1 \cdot \nrd_2 \vecsigma_1 \cdot
\vecsigma_2  [ {\bf 1}, {\bm
    \tau}_1\cdot {\bm \tau}_2, {\bm \tau}_1\cdot {\bm \tau}_3]$
&
$o_{79-82}$ & $i \nrd_1\times\nr_2\cdot\vecsigma_2 
  [ {\bf 1}, {\bm
    \tau}_1\cdot {\bm \tau}_2, {\bm \tau}_1\cdot {\bm \tau}_3,{\bm \tau}_2\cdot {\bm \tau}_3]$
\\
$o_{13-16}$  & $\nrd_1 \cdot \vecsigma_1 \nrd_2 \cdot
\vecsigma_3  [ {\bf 1}, {\bm
    \tau}_1\cdot {\bm \tau}_2, {\bm \tau}_1\cdot {\bm \tau}_3, {\bm
    \tau}_2 \cdot {\bm \tau}_3]$
&
$o_{83-86}$ & $i \nrd_1\times\nr_2\cdot\vecsigma_3 
  [ {\bf 1}, {\bm
    \tau}_1\cdot {\bm \tau}_2, {\bm \tau}_1\cdot {\bm \tau}_3,{\bm \tau}_2\cdot {\bm \tau}_3]$
\\
$o_{17-20}$  & $\nrd_1 \cdot \vecsigma_3 \nrd_2 \cdot
\vecsigma_1  [ {\bf 1}, {\bm
    \tau}_1\cdot {\bm \tau}_2, {\bm \tau}_1\cdot {\bm \tau}_3, {\bm
    \tau}_2 \cdot {\bm \tau}_3]$
&
$o_{87-90}$ & $i \nrd_1\cdot \nr_2 \vecsigma_1\times\vecsigma_2\cdot\vecsigma_3 
  [ {\bf 1}, {\bm
    \tau}_1\cdot {\bm \tau}_2, {\bm \tau}_1\cdot {\bm \tau}_3,{\bm \tau}_2\cdot {\bm \tau}_3]$
\\
$o_{21-24}$  & $\nrd_1 \cdot \nrd_2 \vecsigma_1 \cdot
\vecsigma_3  [ {\bf 1}, {\bm
    \tau}_1\cdot {\bm \tau}_2, {\bm \tau}_1\cdot {\bm \tau}_3, {\bm
    \tau}_2 \cdot {\bm \tau}_3]$
&
$o_{91-94}$ & $i \nrd_1\cdot  \vecsigma_1 \nr_2 \times\vecsigma_2\cdot\vecsigma_3 
  [ {\bf 1}, {\bm
    \tau}_1\cdot {\bm \tau}_2, {\bm \tau}_1\cdot {\bm \tau}_3,{\bm \tau}_2\cdot {\bm \tau}_3]$
\\
$o_{25}$     & $\nrd_1 \times \nrd_2 \cdot \vecsigma_1
[ {\bm \tau}_1 \times {\bm \tau}_2 \cdot {\bm \tau}_3]$ 
&
$o_{95-98}$ & $i \nrd_1\cdot  \vecsigma_2 \nr_2 \times\vecsigma_1\cdot\vecsigma_3 
  [ {\bf 1}, {\bm
    \tau}_1\cdot {\bm \tau}_2, {\bm \tau}_1\cdot {\bm \tau}_3,{\bm \tau}_2\cdot {\bm \tau}_3]$
\\
$o_{26}$     & $\nrd_1 \times \nrd_2 \cdot \vecsigma_3
[ {\bm \tau}_1 \times {\bm \tau}_2 \cdot {\bm \tau}_3]$ 
&
$o_{99-102}$ & $i \nrd_1\cdot  \vecsigma_3 \nr_2 \times\vecsigma_1\cdot\vecsigma_2 
  [ {\bf 1}, {\bm
    \tau}_1\cdot {\bm \tau}_2, {\bm \tau}_1\cdot {\bm \tau}_3,{\bm \tau}_2\cdot {\bm \tau}_3]$
\\
$o_{27}$     & $\nrd_1 \cdot \nrd_2 \vecsigma_1 \times \vecsigma_2 \cdot \vecsigma_3
 [{\bm \tau}_1 \times {\bm \tau}_2 \cdot {\bm \tau}_3]$ 
&
$o_{103-106}$ & $i \nrd_1 \times\vecsigma_2\cdot\vecsigma_3 \nr_2\cdot
\vecsigma_1   
  [ {\bf 1}, {\bm
    \tau}_1\cdot {\bm \tau}_2, {\bm \tau}_1\cdot {\bm \tau}_3,{\bm \tau}_2\cdot {\bm \tau}_3]$
\\
$o_{28}$     & $\nrd_1 \cdot \vecsigma_1 \vecsigma_2 \times \vecsigma_3 \cdot \nrd_2
[ {\bm \tau}_1 \times {\bm \tau}_2 \cdot {\bm \tau}_3]$ 
&
$o_{107-110}$ & $i \nrd_1 \times\vecsigma_1\cdot\vecsigma_3 \nr_2\cdot
\vecsigma_2   
  [ {\bf 1}, {\bm
    \tau}_1\cdot {\bm \tau}_2, {\bm \tau}_1\cdot {\bm \tau}_3,{\bm \tau}_2\cdot {\bm \tau}_3]$
\\
$o_{29}$     & $\nrd_1 \cdot \vecsigma_2 \vecsigma_1 \times \vecsigma_3 \cdot \nrd_2
[ {\bm \tau}_1 \times {\bm \tau}_2 \cdot {\bm \tau}_3]$ 
&
$o_{111-114}$ & $i \nrd_1 \times\vecsigma_1\cdot\vecsigma_2 \nr_2\cdot
\vecsigma_3   
  [ {\bf 1}, {\bm
    \tau}_1\cdot {\bm \tau}_2, {\bm \tau}_1\cdot {\bm \tau}_3,{\bm \tau}_2\cdot {\bm \tau}_3]$
\\
$o_{30}$     & $\nrd_1 \cdot \vecsigma_3 \vecsigma_1 \times \vecsigma_2 \cdot \nrd_2
[ {\bm \tau}_1 \times {\bm \tau}_2 \cdot {\bm \tau}_3]$ 
&
$o_{115-118}$ & $i \nrd_1 \times \nr_2 \cdot \vecsigma_1 \vecsigma_2
\cdot \vecsigma_3   
  [ {\bf 1}, {\bm
    \tau}_1\cdot {\bm \tau}_2, {\bm \tau}_1\cdot {\bm \tau}_3,{\bm \tau}_2\cdot {\bm \tau}_3]$
\\
$o_{31}$     & $\nrd_1 \times \nrd_2 \cdot \vecsigma_1  \vecsigma_2 \cdot \vecsigma_3
[ {\bm \tau}_1 \times {\bm \tau}_2 \cdot {\bm \tau}_3]$ 
&
$o_{119-122}$ & $i \nrd_1 \times \nr_2 \cdot \vecsigma_2 \vecsigma_1
\cdot \vecsigma_3   
  [ {\bf 1}, {\bm
    \tau}_1\cdot {\bm \tau}_2, {\bm \tau}_1\cdot {\bm \tau}_3,{\bm \tau}_2\cdot {\bm \tau}_3]$
\\
$o_{32}$     &  $\nrd_1 \times \nrd_2 \cdot \vecsigma_3  \vecsigma_1 \cdot \vecsigma_2
[ {\bm \tau}_1 \times {\bm \tau}_2 \cdot {\bm \tau}_3]$ 
&
$o_{123-126}$ & $i \nrd_1 \times \nr_2 \cdot \vecsigma_3 \vecsigma_1
\cdot \vecsigma_2   
  [ {\bf 1}, {\bm
    \tau}_1\cdot {\bm \tau}_2, {\bm \tau}_1\cdot {\bm \tau}_3,{\bm \tau}_2\cdot {\bm \tau}_3]$
\\
$o_{33-64}$ & same as $o_{1-32}$  with $\nrd \to \nr$
&
$o_{127-129}$ & $\nrd_1\cdot \nrd_1
 [ {\bf 1}, {\bm
    \tau}_1\cdot {\bm \tau}_2,{\bm \tau}_2\cdot {\bm \tau}_3]$
\\
$o_{65}$ & $i \nrd_1 \cdot \nr_2 
[ {\bm \tau}_1 \times {\bm \tau}_2 \cdot {\bm \tau}_3]$ 
&
$o_{130-133}$ & $\nrd_1\cdot \nrd_1 \vecsigma_1 \cdot \vecsigma_2
 [ {\bf 1}, {\bm
    \tau}_1\cdot {\bm \tau}_2,{\bm \tau}_2\cdot {\bm \tau}_3,{\bm
    \tau}_1\cdot {\bm \tau}_3]$ 
\\
$o_{66}$ & $i \nrd_1 \cdot \vecsigma_1 \nr_2 \cdot \vecsigma_2
[ {\bm \tau}_1 \times {\bm \tau}_2 \cdot {\bm \tau}_3]$ 
&
$o_{134-136}$ & $\nrd_1\cdot \nrd_1 \vecsigma_2 \cdot \vecsigma_3
 [ {\bf 1}, {\bm
    \tau}_1\cdot {\bm \tau}_2,{\bm \tau}_2\cdot {\bm \tau}_3]$ 
\\
$o_{67}$ & $i \nrd_1 \cdot \vecsigma_2 \nr_2 \cdot \vecsigma_1
[ {\bm \tau}_1 \times {\bm \tau}_2 \cdot {\bm \tau}_3]$ 
&
$o_{137-140}$ & $\nrd_1\cdot \vecsigma_1 \nrd_1 \cdot \vecsigma_2 
 [ {\bf 1}, {\bm
    \tau}_1\cdot {\bm \tau}_2,{\bm \tau}_2\cdot {\bm \tau}_3,{\bm
    \tau}_1\cdot {\bm \tau}_3]$ 
\\
$o_{68}$ & $i \nrd_1 \cdot \nr_2 \vecsigma_1\cdot \vecsigma_2
[ {\bm \tau}_1 \times {\bm \tau}_2 \cdot {\bm \tau}_3]$ 
&
$o_{141-143}$ & $\nrd_1\cdot \vecsigma_2 \nrd_1 \cdot \vecsigma_3 
 [ {\bf 1}, {\bm
    \tau}_1\cdot {\bm \tau}_2,{\bm \tau}_2\cdot {\bm \tau}_3]$ 
\\
$o_{69}$ & $i \nrd_1 \cdot \vecsigma_1 \nr_2 \cdot \vecsigma_3
[ {\bm \tau}_1 \times {\bm \tau}_2 \cdot {\bm \tau}_3]$ 
&
$o_{144}$ & $\nrd_1\cdot  \nrd_1 \vecsigma_1 \times \vecsigma_2 \cdot \vecsigma_3 
[ {\bm \tau}_1 \times {\bm \tau}_2 \cdot {\bm \tau}_3]$ 
\\
$o_{70}$ & $i \nrd_1 \cdot \vecsigma_3 \nr_2 \cdot \vecsigma_1
[ {\bm \tau}_1 \times {\bm \tau}_2 \cdot {\bm \tau}_3]$ 
&
$o_{145}$ & $\nrd_1\cdot   \vecsigma_1 \nrd_1 \times \vecsigma_2 \cdot \vecsigma_3 [ {\bm \tau}_1 \times {\bm \tau}_2 \cdot {\bm \tau}_3]$ 
\\
$o_{71}$ & $i \nrd_1 \cdot \nr_2 \vecsigma_1\cdot \vecsigma_3
[ {\bm \tau}_1 \times {\bm \tau}_2 \cdot {\bm \tau}_3]$ 
&
$o_{146}$ & $\nrd_1\cdot   \vecsigma_2 \nrd_1 \times \vecsigma_1 \cdot \vecsigma_3[ {\bm \tau}_1 \times {\bm \tau}_2 \cdot {\bm \tau}_3]$ 
\\
$o_{72}$ & $i \nrd_1 \cdot \vecsigma_2 \nr_2 \cdot \vecsigma_3
[ {\bm \tau}_1 \times {\bm \tau}_2 \cdot {\bm \tau}_3]$ 
&&\\
\hline
\end{tabular}
\end{scriptsize}
\caption{Complete list of 2-derivative three-nucleon contact operators
  compatible with rotational, isospin, parity and time-reversal
  invariance. Subscripts refer to the nucleon bilinear on which the
  operators act. For instance $o_{73}=\epsilon^{abc} (N^\dagger \nrd_i
  \tau^a N) \nabla_j ( N^\dagger \sigma^j \tau^b N) N^\dagger \sigma^i
  \tau^c N$.\label{tab:operators}}
\end{table}
Fierz rearrangements also involve the field derivatives: for instance, under
permutation of nucleons 1-2,
\begin{equation}
\nrd_1 \to \frac{1}{2} ( \nr_2 + \nrd_2 -\nr_1 + \nrd_1), \quad
\nr_1  \to \frac{1}{2} ( \nr_2 + \nrd_2 +\nr_1 - \nrd_1).
\end{equation}
Out of the $3\times 146$ relations from the Fierz identities, 128 are linearly independent, and displayed in the Appendix~\ref{app:rel}. There remain 18 independent operators, which can be chosen to be
\begin{equation}
\label{eq:14operators}
o_1,\quad o_2, \quad o_4,\quad o_5, \quad o_7, \quad
o_{33},\quad o_{34},\quad o_{35},\quad o_{36},\quad o_{37},\quad o_{38},\quad o_{39},\quad
o_{40},\quad o_{42},\quad o_{43},\quad o_{45},\quad o_{58},\quad o_{64}.
\end{equation}

\section{Further constraints from relativity} \label{sec:rel}
We still have to impose the requirements of Poincar\'e covariance. They can be
implemented order by order in the low-energy expansion, as detailed in
Ref.~\cite{relativity}. As a result, the subleading $3N$
effective Hamiltonian consists of a set of terms whose strength is
fixed by the lowest order coupling constant, and a set of free terms
which have to commute with the lowest order free boost operator. In the
following we focus our attention on the latter set.
With the choice for the non-relativistic nucleon field $N(x)$,
\begin{equation}
N(x) = \int \frac{d{\bf p}}{(2\pi)^3}\, b_s({\bf  p}) \, \chi_s\, {\rm e}^{-i p \cdot x} \ ,
\label{eq:nrf}
\end{equation}
where $b_s({\bf p})$ and $b^\dagger_s({\bf p})$
are annihilation and creation operators for a nucleon in spin state $s$,
satisfying standard anticommutation relations, {\it i.e.}
$\left[b_s({\bf p})\, , \, b^\dagger_{s^\prime }({\bf p}^\prime)\right]_+ = (2\pi)^3\delta({\bf p}-{\bf p}^\prime)\,
\delta_{ss^\prime}$,
the leading order free boost operators ${\bf K}_0$ acts in the following way
\cite{relativity}
\begin{equation} \label{eq:lcomm}
 \left[ {\bf K}_0 \, , \, b_s ({\bf p} )\right] = -i \,m\,
{\bm \nabla}_{\bf p} \, b_s ({\bf p}) \ .
\end{equation}
Only the first 5, out of the 18 operators listed in
Eq.~(\ref{eq:14operators}), do not commute with ${\bf K}_0$
\begin{equation}
\label{eq:commutators}
\begin{array}{lll}
\left[{\bf K}_0,o_1\right] = -4 m \nrd_1, &
\left[{\bf K}_0,o_2\right] = -4 m \nrd_1 {\bm \tau}_1 \cdot {\bm \tau}_2, &
\left[{\bf K}_0,o_4\right] = -4 m \nrd_1\cdot \vecsigma_1 \vecsigma_2,  \\
\left[{\bf K}_0,o_5\right] = -4 m \nrd_1\cdot \vecsigma_1 \vecsigma_2 {\bm \tau}_1
\cdot {\bm \tau}_2, &\left[{\bf K}_0,o_7\right] = -4 m \nrd_1\cdot \vecsigma_2 \vecsigma_1,
\end{array}
\end{equation}
in the same notations as in Table~\ref{tab:operators}.
The vector operators in the right hand sides of
Eq.~(\ref{eq:commutators}), in turn, mix with 44 others under Fierz
rearrangements; for instance, by exchanging the indeces of particles 1-2 we get the identity 
\begin{eqnarray}
\nrd_1 \cdot \vecsigma_1 \vecsigma_2 &=& \frac{1}{4} \nrd_1 + \frac{1}{4} \nrd_1 {\bm \tau}_1 \cdot {\bm \tau}_2 - \frac{1}{4} \nrd_1 \vecsigma_1\cdot \vecsigma_2 -\frac{1}{4} \nrd_1 \vecsigma_1\cdot \vecsigma_2 {\bm \tau}_1\cdot {\bm \tau}_2 \nonumber \\
&& +\frac{1}{4} \nrd_1 \cdot \vecsigma_1 \vecsigma_2 + \frac{1}{4} \nrd_1 \cdot \vecsigma_1 \vecsigma_2 {\bm \tau}_1 \cdot {\bm \tau}_2 + \frac{1}{4} \nrd_1 \cdot \vecsigma_2 \vecsigma_1   + \frac{1}{4} \nrd_1 \cdot \vecsigma_2 \vecsigma_1 {\bm \tau}_1 \cdot {\bm \tau}_2
\nonumber \\
&& -\frac{3}{4} i \nr_1 \times \vecsigma_2  - \frac{1}{2} i \nr_1 \times \vecsigma_2 {\bm \tau}_1 \cdot {\bm \tau}_2 - \frac{1}{4} i \nr_1 \times \vecsigma_2 {\bm \tau}_2 \cdot {\bm \tau}_3.
\end{eqnarray} 
By analyzing all the 3$\times$49 relations we find that the five operators on the right hand sides of Eq.~(\ref{eq:commutators}) 
 are linearly independent and, in fact, form a basis of
all vector operators satisfying the symmetry requirements. Therefore
we conclude that the operators $o_{1,2,4,5,7}$ 
are forbidden by Poincar\'e symmetry, and we are left with the
remaining 13 operators.
The same conclusion is reached by starting the analysis with an
operator basis written in terms of gradients with respect to relative coordinates (or Jacobi momenta, in conjugate space),  
\begin{equation}
\nabla_a  \propto \nabla_2 - \nabla_1, \quad \nabla_b \propto \nabla_3 - \frac{1}{2} \left( \nabla_1 + \nabla_2\right),
\end{equation}
in which case one can write an initial list of 116 operators and reduce their number to 13 using the Fierz's constraints.
This in turns provides a non-trivial check of our calculation. A
minimal form of the 2-derivative $3N$ contact Lagrangian is given by
\begin{eqnarray}
{\cal L}^{(2)}_{3N} &=& 
 E^\prime_1 \nr (N^\dagger N) \cdot \nr (N^\dagger N) N^\dagger N   
+ E^\prime_2 \nr (N^\dagger \tau^a N) \cdot \nr (N^\dagger \tau^a N)
N^\dagger N   \nonumber \\
&&+ E^\prime_3 \nr (N^\dagger \tau^a N) \cdot \nr (N^\dagger  N) N^\dagger
\tau^a N
+ E^\prime_4 \nr \cdot (N^\dagger \vecsigma N)  \nr \cdot (N^\dagger
\vecsigma N) N^\dagger N \nonumber \\
&&+ E^\prime_5 \nr \cdot (N^\dagger \vecsigma \tau^a N)  \nr \cdot (N^\dagger
\vecsigma \tau^a N) N^\dagger N
+E_6^\prime\nr \cdot (N^\dagger \vecsigma \tau^a N)  \nr \cdot (N^\dagger
\vecsigma  N) N^\dagger \tau^a N \nonumber \\
&& + E^\prime_7 \nr_i (N^\dagger \sigma^j   N)  \nr_j  (N^\dagger
\sigma^i  N) N^\dagger N
+ E^\prime_8 \nr_i (N^\dagger \sigma^j  \tau^a N)  \nr_j  (N^\dagger
\sigma^i  \tau^a N) N^\dagger N 
\nonumber \\
&&+ E^\prime_{9} \nr_i (N^\dagger \sigma^j  N)  \nr_i  (N^\dagger
\sigma^j  N) N^\dagger N
+ E^\prime_{10} \nr_i (N^\dagger \sigma^j \tau^a N)  \nr_i  (N^\dagger
\sigma^j \tau^a N) N^\dagger N \nonumber \\
&& + E^\prime_{11} \nr \cdot (N^\dagger \vecsigma N)  \nr  (N^\dagger
 N) \cdot N^\dagger \vecsigma N 
+ E^\prime_{12} \epsilon^{abc} \nr  (N^\dagger \tau^a  N)  \times \nr
(N^\dagger  \tau^b N) \cdot
N^\dagger \vecsigma \tau^c N \nonumber \\
&&+ E^\prime_{13} \epsilon^{abc} \nr  (N^\dagger \sigma^i \tau^a  N)  \times \nr
(N^\dagger \sigma^i \tau^b N) \cdot
N^\dagger \vecsigma \tau^c N .
\end{eqnarray}

\section{The subleading three-nucleon contact potential} \label{sec:pot}
The $3N$ potential is obtained by taking the matrix element of the
interaction between $3N$ states. Denoting ${\bf k}_i={\bf p}_i -
{\bf p}^\prime_i$ and ${\bf Q}_i={\bf p}_i + {\bf p}^\prime_i$, ${\bf
  p}_i$ and ${\bf p}^\prime_i$ being the initial and final momenta of
nucleon $i$, the potential in momentum space is found to be
\begin{eqnarray}
V=\sum_{i\neq j\neq k} &&\biggl[ -E_1 {\bf k}_i^2 - E_2 {\bf k}_i^2 {\bm
  \tau}_i\cdot {\bm \tau}_j - E_3{\bf k}_i^2 {\bm
  \sigma}_i\cdot {\bm \sigma}_j - E_4 {\bf k}_i^2 {\bm
  \sigma}_i\cdot {\bm \sigma}_j {\bm
  \tau}_i\cdot {\bm \tau}_j - E_5 \left( 3 {\bf k}_i \cdot {\bm \sigma}_i {\bf k}_i\cdot {\bm
    \sigma}_j - {\bf k}_i^2 {\bm \sigma}_i \cdot {\bm \sigma}_j\right) \nonumber \\
  &&-E_6 \left( 3 {\bf k}_i \cdot {\bm \sigma}_i {\bf k}_i\cdot {\bm
  \sigma}_j - {\bf k}_i^2 {\bm \sigma}_i \cdot {\bm \sigma}_j\right) {\bm
  \tau}_i\cdot {\bm \tau}_j
-\frac{i}{4} E_7  {\bf k}_i \times \left({\bf Q}_i - {\bf Q}_j\right) \cdot ({\bm \sigma}_i +
          {\bm \sigma}_j) \nonumber \\
&&-\frac{i}{4} E_8  {\bf k}_i \times \left({\bf Q}_i - {\bf Q}_j\right) \cdot ({\bm \sigma}_i +
          {\bm \sigma}_j) {\bm  \tau}_j\cdot {\bm \tau}_k
 - E_9 {\bf k}_i \cdot {\bm \sigma}_i {\bf k}_j \cdot {\bm \sigma}_j
- E_{10} {\bf k}_i \cdot {\bm \sigma}_i {\bf k}_j \cdot {\bm \sigma}_j
{\bm  \tau}_i\cdot {\bm \tau}_j \nonumber \\
&&- E_{11} {\bf k}_i \cdot {\bm \sigma}_j {\bf k}_j \cdot {\bm \sigma}_i
- E_{12} {\bf k}_i \cdot {\bm \sigma}_j {\bf k}_j \cdot {\bm \sigma}_i
{\bm  \tau}_i\cdot {\bm \tau}_j   - E_{13} {\bf k}_i \cdot {\bm \sigma}_j {\bf k}_j \cdot {\bm \sigma}_i
{\bm  \tau}_i\cdot {\bm \tau}_k \biggr]  , 
\end{eqnarray}
with
\begin{equation}
\begin{array}{l}
  E_1= {1\over 2} E^\prime_1 + {3\over 2 }E^\prime_2 -{3\over 2} E^\prime_3 - {1\over 2} E^\prime_9 + {3\over 2} E^\prime_{10}
,\\
E_2 =  E^\prime_2 + E^\prime_6 +2 E^\prime_{12} - 8 E^\prime_{13},\\
E_3 =  E^\prime_2- E^\prime_3 - E^\prime_6 +{ 1\over 3 } E^\prime_{11} - 2 E^\prime_{12} + 8 E^\prime_{13},\\
E_4 =  {1\over 3 } E^\prime_2 - {1\over 3 } E^\prime_3 -{1\over 3 } E^\prime_6 -{1\over 3 } E^\prime_9 + E^\prime_{10} -{2\over 3 } E^\prime_{12} +{8\over 3 } E^\prime_{13},\\
E_5 =  E^\prime_2 - E^\prime_3 - E^\prime_6 - E^\prime_9 +{1\over 3 } E^\prime_{11} - 2 E^\prime_{12} + 8 E^\prime_{13},\\
E_6 =  {1\over 3 } E^\prime_2 - {1\over 3 } E^\prime_3 - {1\over 3 } E^\prime_6 -{1\over 3 } E^\prime_9 -{2\over 3 } E^\prime_{12} + {8\over 3 } E^\prime_{13},\\
E_7 =  -8 E^\prime_{12} +40 E^\prime_{13},\\
E_8 =  -4 E^\prime_{12} +12 E^\prime_{13},\\
E_9 =  {3\over 2 } E^\prime_2 -{3\over 2 } E^\prime_3 -E^\prime_4 -{3\over 2 } E^\prime_9 +E^\prime_{11} - 3 E^\prime_{12} +12 E^\prime_{13},\\
E_{10} =  {1\over 2 } E^\prime_2 - {1\over 2 } E^\prime_3 - E^\prime_5 -{1\over 2 } E^\prime_9 - E^\prime_{12} +6 E^\prime_{13},\\
E_{11} =  {3\over 2 } E^\prime_2 -{3\over 2 } E^\prime_3 - E^\prime_7 -{3\over 2 } E^\prime_9 + 3 E^\prime_{12} - 12 E^\prime_{13},\\
E_{12} =  {1\over 2 } E^\prime_2 -{1\over 2 } E^\prime_3 - E^\prime_8 - {1\over 2 } E^\prime_9 + E^\prime_{12} - 6 E^\prime_{13},\\
E_{13} =   E^\prime_6 +4 E^\prime_{12} -16 E^\prime_{13}.
\end{array}
\end{equation}

A local form of the $3N$ potential in configuration space can be obtained by using a momentum cutoff depending only on ${\bf k}_i$, as done in Ref.~\cite{navratil}, e.g. $F({\bf k}_j^2; \Lambda) F({\bf k}_k^2; \Lambda)$. In this way the result of the Fourier transform is expressed in terms of the function 
\begin{equation}
Z_0(r;\Lambda)=\int \frac{d {\bf p}}{(2 \pi)^3} {\mathrm{e}}^{i {\bf p}\cdot {\bf r}} F({\bf p}^2;\Lambda)
\end{equation}
and derivatives theoreof. Explicitly, omitting the argument $\Lambda$ in the function $Z_0$,
\begin{eqnarray}
V=\sum_{i\neq j\neq k} && (E_1 + E_2 {\bm \tau}_i \cdot {\bm \tau}_j + E_3 {\bm \sigma}_i \cdot {\bm \sigma}_j + E_4 {\bm \tau}_i \cdot {\bm \tau}_j  {\bm \sigma}_i \cdot {\bm \sigma}_j)  \left[ Z_0^{\prime\prime}(r_{ij}) + 2 \frac{Z_0^\prime(r_{ij})}{r_{ij}}\right] Z_0(r_{ik})  \nonumber \\
&& + (E_5 +E_6 {\bm \tau}_i\cdot{\bm \tau}_j) S_{ij} \left[ Z_0^{\prime\prime}(r_{ij}) - \frac{Z_0^\prime(r_{ij})}{r_{ij}}\right] Z_0(r_{ik})+ (E_7 + E_8 {\bm \tau}_i\cdot{\bm \tau}_k) ( {\bf L}\cdot {\bm S})_{ij} \frac{Z_0^\prime(r_{ij})}{r_{ij}} Z_0(r_{ik}) \nonumber \\
&& +\left[  (E_9 + E_{10} {\bm \tau}_j \cdot {\bm \tau}_k) {\bm \sigma}_j \cdot \hat {\bf r}_{ij}  {\bm \sigma}_k \cdot \hat {\bf r}_{ik}   + (E_{11} + E_{12} {\bm \tau}_j \cdot {\bm \tau}_k+ E_{13} {\bm \tau}_i \cdot {\bm \tau}_j) {\bm \sigma}_k \cdot \hat {\bf r}_{ij}  {\bm \sigma}_j \cdot \hat {\bf r}_{ik} \right] Z_0^\prime(r_{ij}) Z_0^\prime(r_{ik})
\end{eqnarray}
where $S_{ij}$ and $ ( {\bf L}\cdot {\bm S})_{ij}$ are respectively the tensor and spin-orbit operators for particles $i$ and $j$. Notice that a choice of basis has been made such that most terms in the potential can be viewed as an ordinary interaction of particles $ij$ with a further dependence on the coordinate of the third particle. In particular the terms multiplying $E_7$ and $E_8$ are of a spin-orbit character. A specific combination of both has been suggested in Ref.~\cite{kievsky}. We also observe  that some of the spin-isospin structures implied by Eq.~(\ref{eq:lolagr}), which were equivalent up to cutoff effects, are resolved at the two-derivative level. 

\section{Leading four-nucleon Lagrangian} \label{sec:4n}
From the previous discussion, the $3N$ contact interaction consists of a leading contribution,  at the N2LO of the chiral expansion, and a subleading one, arising at the N4LO. Parity requires that no $3N$ contact interactions appear at N3LO or N5LO.
On the other hand,  the leading $4N$ contact interaction arises at N5LO, therefore one is lead to consider such terms at the same level of accuracy, at least in the framework of pionless EFT. By listing all possible spin-isospin structures which respect the discrete symmetries of strong interactions one gets 16 operators, so that the Lagrangian is written as
\begin{eqnarray}
{\cal L}^{(0)}_{\mathrm{4N}} &=& \label{eq:4nlagr}
- F_1 N^\dagger N N^\dagger N N^\dagger N N^\dagger N 
- F_2 N^\dagger \tau^a N N^\dagger \tau^a  N N^\dagger N N^\dagger N  \nonumber \\
&&- F_3 N^\dagger \tau^a N N^\dagger \tau^a  N N^\dagger \tau^b N N^\dagger \tau^b N  
- F_4 N^\dagger \sigma^i N N^\dagger \sigma^i  N N^\dagger N N^\dagger N \nonumber \\
&&- F_5 N^\dagger \sigma^i \tau^a N N^\dagger \sigma^i  \tau^a N N^\dagger N N^\dagger N
- F_6 N^\dagger \sigma^i \tau^a N N^\dagger \sigma^i  N N^\dagger \tau^a N N^\dagger N
\nonumber \\
&& 
- F_7 N^\dagger \sigma^i N N^\dagger \sigma^i  N N^\dagger \tau^a N N^\dagger \tau^a N
- F_8 N^\dagger \sigma^i \tau^a N N^\dagger \sigma^i \tau^a  N N^\dagger \tau^b N N^\dagger \tau^b N
\nonumber \\
&& 
- F_9 N^\dagger \sigma^i \tau^a N N^\dagger \sigma^i \tau^b  N N^\dagger \tau^a N N^\dagger \tau^b N
- F_{10} N^\dagger \sigma^i N N^\dagger \sigma^i  N N^\dagger \sigma^j N N^\dagger \sigma^j N
\nonumber \\
&& 
- F_{11} N^\dagger \sigma^i \tau^a N N^\dagger \sigma^i \tau^a N N^\dagger \sigma^j N N^\dagger \sigma^j N
- F_{12} N^\dagger \sigma^i \tau^a N N^\dagger \sigma^i  N N^\dagger \sigma^j \tau^a N N^\dagger \sigma^j N
\nonumber \\
&& 
- F_{13} N^\dagger \sigma^i \tau^a N N^\dagger \sigma^i \tau^a N N^\dagger \sigma^j \tau^b N  N^\dagger \sigma^j  \tau^b N
- F_{14} N^\dagger \sigma^i \tau^a N N^\dagger \sigma^i \tau^b N N^\dagger \sigma^j \tau^a N  N^\dagger \sigma^j  \tau^b N
\nonumber \\
&& 
- F_{15} \epsilon^{ijk} \epsilon^{abc} N^\dagger \sigma^i \tau^a N N^\dagger \sigma^j \tau^b N N^\dagger \sigma^k \tau^c N  N^\dagger  N
- F_{16} \epsilon^{ijk} \epsilon^{abc} N^\dagger \sigma^i \tau^a N N^\dagger \sigma^j \tau^b N N^\dagger \sigma^k  N  N^\dagger \tau^c N
\nonumber \\
&\equiv& -\sum_{i=1}^{16} F_i O_i^{4N}.
\end{eqnarray}
 However, after using all possible Fierz rearrangements (cfr. Appendix~\ref{app:rel}), as was done in the $3N$ sector, it turns out that they are all equivalent, leaving only one independent $4N$ contact operator. Using cutoff functions depending only on ${\bf k}_i$, e.g. choosing $F({\bf k}_j^2; \Lambda) F({\bf k}_k^2; \Lambda) F({\bf k}_l^2; \Lambda)$, the 4-body contact potential is local in coordinate space,
\begin{equation}
V_{4N}=F \sum_{i\neq j\neq k\neq l} Z_0(r_{ij}) Z_0(r_{ik}) Z_0 (r_{il}),
\end{equation}
 where $F$ is the single accompanying LEC.
This allows to decouple, to a certain extent, the $3N$ sector from the $4N$ sector, if one is willing to fix all the LECs from the data. For instance, one can in principle adjust the subleading $3N$ potential in the $A=3$ systems without worrying much about the consequences for the $\alpha$ particle binding energy: the latter could always be reproduced by adjusting the $4N$ contact term. 

\section{Conclusions}
In the present paper we have derived a $3N$ potential
from the  minimal form of the two-derivative $3N$ contact
Lagrangian. This potential has 13 unknown LECs and,
with a particular choice of the cutoff, can be put in a local form.
In addition we have shown that the leading $4N$ contact Lagrangian consists of only one operartor.

It should be stressed that these terms start to contribute at N4LO  and are therefore beyond the accuracy of the presently developed nuclear interactions. In particular, a complete EFT calculation should also address the problem of the $NN$ interaction at the same order. Notice that the four-derivatives $2N$ operators are already part of the N3LO interaction while the six-derivative ones would start to contribute at N5LO, and have not been considered in the literature so far. 
Nevertheless, despite respresenting only part of the N4LO interaction, the terms derived here could play an important role in the accurate description of the three-nucleon systems, since they are completely unconstrained by symmetries. This is at variance with the $3N$ N3LO interaction, which contains no free parameters \cite{epelbaum08,bernard08}.
Moreover, the same terms will also appear in the pionless version of the effective theory as next-to-leading $3N$ interaction and leading $4N$  interaction.

The utility of the derived potential could be the following:
After a sensitivity study of the 13 subleading terms of the $3N$
potentials, the corresponding LECs can be
determined from a fit of several polarization observables in
N-d scattering at low energies. There are well established
discrepancies between theoretical predictions and experimental
data in some of the polarization observables as for example the
vector analyzing powers $A_y$ and $iT_{11}$ and the vector
analyzing power $T_{21}$ in elastic scattering \cite{koike,kievsky01,sagara}, and in several unpolarized breakup cross sections \cite{glockle,sagara}. Our expectation,
based on the results of Ref.~\cite{kievsky}, is that some of the 13 operators of
the $3N$ potential will have sufficient sensitivity to
fix these discrepancies with reasonable values of their LECs.
At the same time the 4N contact term will be used to reproduce the $^4$He binding
energy. Finally, the predictions of the derived potentials will be
tested in the description of the four-body scattering states. In fact, also in this case there exist large and still unexplained discrepancies between theory and experiment \cite{fisher,deltuva1,deltuva2,deltuva3}. Studies along these lines are in progress.

\appendix

\section{Linear relations from Fierz's constraints} \label{app:rel}
In this Appendix we list the linearly independent relations among the 146 operators of Table~\ref{tab:operators}. The following ones are obtained by applying Fierz's reshuffling of indeces of particles 1-2:

{\allowdisplaybreaks 
\begin{align} 
2 o_{25} =& o_{76} + 2 o_{77} - 3
o_{78} + 2 o_{81} - 2 o_{82} + o_{84} - o_{86} \\
2  o_{26} =&o_{80} -
o_{82} + 2 o_{85} - 2 o_{86}+o_{120} -
o_{122} + 2 o_{125} - 2 o_{126}  \\
 2 o_{32} =&3 o_{80} - 3 o_{82} + 6
o_{85} - 6 o_{86} - o_{120} + o_{122} - 2 o_{125}
                                                                     + 2 o_{126} \\
8 o_{35} =&- 2 o_{3} - 2 o_{12} + o_{34} + o_{35} + o_{54} + o_{56} + o_{128} + o_{129}
                                                              + o_{132} + o_{133} \\
2 o_{57} =&- o_{76} + o_{78} - o_{84} + o_{86} \\
4 o_{62} =& 2 o_{15} - 2 o_{16} + 2 o_{19} - 2 o_{20} - o_{37} + o_{38} - o_{40} + o_{41}
    + o_{46} - o_{47} + o_{50} - o_{51} - 2 o_{138} \nonumber \\
&+ 2 o_{139} + 2 o_{142} - 2 o_{143} \\
2 o_{63} =& - o_{80} + o_{82} - o_{116} + o_{118} + o_{120} - o_{122} - o_{124} + o_{126} \\
2 o_{64} =&  - 3 o_{80} + 3 o_{82} + o_{120} - o_{122} \\
4 o_{65} =&o_{34} - o_{35} + o_{54} - o_{56} + o_{128} - o_{129} - o_{132} + o_{133} \\
4 o_{68} =&3 o_{34} - 3 o_{35} - o_{54} + o_{56} + 3 o_{128} - 3 o_{129} + o_{132}
                                                                       - o_{133} \\
4 o_{87} =&o_{42} + o_{44} - o_{53} - o_{56} + o_{130} + o_{133} - o_{134} - o_{135} \\
4 o_{88} =&3 o_{42} - o_{44} - 3 o_{53} + o_{56} + 3 o_{130} - o_{133} - 3 o_{134}
                                                                       + o_{135} \\
4 o_{123} =&  3 o_{79} + 3 o_{81} + 3 o_{83} + 3 o_{84} - o_{119} - o_{121} - o_{123}
                                                                       - o_{124} \\
4 o_{124} =& 9 o_{79} - 3 o_{81} + 9 o_{83} - 3 o_{84} - 3 o_{119} + o_{121} - 3 o_{123}
                                                                       + o_{124} \\
8 o_{125} =& 3 o_{26} - o_{32} + 3 o_{58} - o_{64} + 6 o_{80} + 6 o_{82} + 6 o_{85}
                          + 6 o_{86} - 2 o_{120} - 2 o_{122} - 2 o_{125} - 2 o_{126},
\end{align}
}
while Fierzing on particles 1-3 we get
{\allowdisplaybreaks
\begin{align}
4 o_1 =& - o_{1} - o_{3} - o_{21} - o_{23} \\
4 o_2 =&- o_{2} - o_{3} - o_{22} - o_{24} - o_{65} - o_{74} \\
4 o_3 =&- 3 o_{1} + o_{3} - 3 o_{21} + o_{23} \\
4 o_4 =&- o_{4} - o_{6} - o_{13} - o_{16} - o_{91} - o_{94} \\
4 o_5 =&- o_{5} - o_{6} - o_{14} - o_{15} + o_{28} - o_{66} - o_{69} - o_{92} - o_{93} \\
4 o_7 =&- o_{7} - o_{9} - o_{17} - o_{20} - o_{103} - o_{106} \\
4 o_8 =& - o_{8} - o_{9} - o_{18} - o_{19} - o_{29} - o_{67} - o_{70} - o_{104} - o_{105}
 \\
4 o_{10} =&         - o_{10} - o_{12} - o_{21} - o_{24} - o_{87} - o_{90} \\
4 o_{13} =& - o_{1} - o_{3} - o_{13} - o_{15} - o_{17} - o_{19} + o_{21} + o_{23} - o_{79}
- o_{82} + o_{83} + o_{86} \\
4 o_{19} =& -3 o_1 + o_3 - 3 o_{13} + o_{15} - 3 o_{17} + o_{19} + 3 o_{21} - o_{23} + 3 o_{79} - o_{82} - 3 o_{83} + o_{86} \\
2 o_{25} =&  o_{14} - o_{16} - o_{18} + o_{20} + o_{80} - o_{81} + o_{84} - o_{85} \\
2 o_{26} =& - o_{14} + o_{16} + o_{18} - o_{20} + o_{80} - o_{81} + o_{84} - o_{85} \\
o_{27} =&          o_{11} - o_{12} - o_{22} + o_{23} \\
2 o_{28} =&- o_{8} + o_{9} + o_{11} - o_{12} + o_{18} - o_{19} - o_{22} + o_{23} + o_{76}
                  - o_{77} - o_{108} + o_{109} - o_{112} \nonumber \\
&+ o_{113} - o_{116} + o_{117} \\
o_{29} =&      - o_{8} + o_{9} + o_{18} - o_{19} \\
2 o_{30} =&- o_{8} + o_{9} + o_{11} - o_{12} + o_{18} - o_{19} - o_{22} + o_{23} - o_{76}
                  + o_{77} + o_{108} - o_{109} + o_{112} \nonumber \\
&- o_{113} + o_{116} - o_{117} \\
2 o_{31} =& o_{5} - o_{6} - o_{8} + o_{9} - o_{14} + o_{15} + o_{18} - o_{19} + o_{76}
                  - o_{77} - o_{116} + o_{117} + o_{120} - o_{121} \nonumber \\
&+ o_{124} - o_{125} \\
4 o_{33} =&                 - o_{33} - o_{35} - o_{53} - o_{55} \\
4 o_{34} =&   - o_{34} - o_{35} - o_{54} - o_{56} + o_{65} + o_{71} \\
4 o_{35} =&           - 3 o_{33} + o_{35} - 3 o_{53} + o_{55} \\
4 o_{36} =&    - o_{36} - o_{38} - o_{45} - o_{48} - o_{107} - o_{109} \\
4 o_{37} =& - o_{37} - o_{38} - o_{46} - o_{47} + o_{60} + o_{66} + o_{73} - o_{108}
                                                                       - o_{110} \\
4 o_{39} =&      - o_{39} - o_{41} - o_{49} - o_{52} - o_{95} - o_{97} \\
4 o_{40} =& - o_{40} - o_{41} - o_{50} - o_{51} - o_{61} + o_{67} + o_{72} - o_{96} - o_{98} \\
4 o_{42} =&     - o_{42} - o_{44} - o_{53} - o_{56} + o_{87} + o_{89} \\
4 o_{45} =& - o_{33} - o_{35} - o_{45} - o_{47} - o_{49} - o_{51} + o_{53} + o_{55} + o_{75}
                                                        + o_{77} - o_{83} - o_{85} \\
2 o_{57} =&  o_{46} - o_{48} - o_{50} + o_{52} - o_{76} + o_{78} - o_{84} + o_{86} \\
o_{59} =&             o_{43} - o_{44} - o_{54} + o_{55} \\
2 o_{60} =& - o_{40} + o_{41} + o_{43} - o_{44} + o_{50} - o_{51} - o_{54} + o_{55} - o_{80}
                    + o_{82} - o_{92} + o_{94} + o_{100} \nonumber \\
&- o_{102} + o_{120} - o_{122} \\
o_{61} =&       - o_{40} + o_{41} + o_{50} - o_{51} \\
2 o_{62} =& - o_{40} + o_{41} + o_{43} - o_{44} + o_{50} - o_{51} - o_{54} + o_{55} + o_{80}
                    - o_{82} + o_{92} - o_{94} - o_{100} \nonumber \\
&+ o_{102} - o_{120} + o_{122}
 \\
2 o_{65} =&      o_{34} - o_{35} + o_{54} - o_{56} \\
2 o_{66} =&  o_{37} - o_{38} + o_{46} - o_{47} + o_{108} - o_{110} \\
2 o_{67} =&    o_{40} - o_{41} + o_{50} - o_{51} + o_{96} - o_{98} \\
2 o_{68} =&     o_{43} - o_{44} + o_{54} - o_{55} - o_{88} + o_{90} \\
2 o_{69} =&o_{34} - o_{35} + o_{46} - o_{48} + o_{50} - o_{52} - o_{54} + o_{56} - o_{76}
                                                        + o_{78} + o_{84} - o_{86}
 \\
2 o_{70} =&o_{34} - o_{35} + o_{46} - o_{48} + o_{50} - o_{52} - o_{54} + o_{56} + o_{76}
                                                        - o_{78} - o_{84} + o_{86} \\
2 o_{71} =&        3 o_{34} - 3 o_{35} - o_{54} + o_{56} \\
2 o_{72} =&   o_{40} - o_{41} + o_{50} - o_{51} - o_{96} + o_{98} \\
2 o_{73} =&   o_{37} - o_{38} + o_{46} - o_{47} - o_{108} + o_{110} \\
2 o_{74} =&    o_{43} - o_{44} + o_{54} - o_{55} + o_{88} - o_{90} \\
4 o_{75} =& o_{45} + o_{47} - o_{49} - o_{51} - o_{75} - o_{77} - o_{83} - o_{85} \\
4 o_{76} =& o_{46} + o_{48} - o_{50} - o_{52} - o_{57} - o_{58} - o_{69} + o_{70} - o_{76}
                                                        - o_{78} - o_{84} - o_{86} \\
4 o_{77} =& 3 o_{45} - o_{47} - 3 o_{49} + o_{51} - 3 o_{75} + o_{77} - 3 o_{83} + o_{85} \\
4 o_{80} =& 2 o_{80} + 2 o_{82}   - o_{58} -  o_{64} \\
4 o_{83} =&- o_{45} - o_{47} + o_{49} + o_{51} - o_{75} - o_{77} - o_{83} - o_{85} \\
4 o_{84} =& - o_{46}- o_{48} + o_{50}+ o_{52} - o_{57}- o_{58}+ o_{69}- o_{70}- o_{76} - o_{78} - o_{84}- o_{86} \\
2 o_{87} =&         o_{42} + o_{44} - o_{53} - o_{56} \\
2 o_{88} =&          o_{43} + o_{44} - o_{54} - o_{55} - o_{68} + o_{74} \\
2 o_{89} =&  3 o_{42} - o_{44} - 3 o_{53} + o_{56} \\
4 o_{91} =& - o_{39} - o_{41} + o_{42} + o_{44} + o_{49} + o_{52} - o_{53} - o_{56} - o_{79}
                     - o_{81} - o_{91} - o_{93} + o_{99} \nonumber \\
&+ o_{101} + o_{119} + o_{121} \\
4 o_{93} =& - 3 o_{39} + o_{41} + 3 o_{42} - o_{44} + 3 o_{49} - o_{52} - 3 o_{53} + o_{56}
     - 3 o_{79} + o_{81} - 3 o_{91} + o_{93}  \nonumber \\
&+ 3 o_{99} - o_{101} + 3 o_{119} - o_{121} \\
2 o_{95} =&   - o_{39} - o_{41} + o_{49} + o_{52} \\
2 o_{96} =&  - o_{40} - o_{41} + o_{50} + o_{51} + o_{67} - o_{72} \\
2 o_{97} =&          - 3 o_{39} + o_{41} + 3 o_{49} - o_{52} \\
4 o_{99} =&- o_{39} - o_{41} + o_{42} + o_{44} + o_{49} + o_{52} - o_{53} - o_{56} + o_{79}
                     + o_{81} + o_{91} + o_{93} - o_{99} \nonumber \\
&- o_{101} - o_{119} - o_{121} \\
4 o_{103} =& - o_{36} - o_{38} + o_{42} + o_{44} + o_{45} + o_{48} - o_{53} - o_{56} + o_{79}
                  + o_{81} - o_{103} - o_{105} + o_{111} \nonumber \\
&+ o_{113} - o_{119} - o_{121} \\
4 o_{105} =& - 3 o_{36} + o_{38} + 3 o_{42} - o_{44} + 3 o_{45} - o_{48} - 3 o_{53} + o_{56}
  + 3 o_{79} - o_{81} - 3 o_{103} + o_{105} \nonumber \\
& + 3 o_{111} - o_{113} - 3 o_{119} + o_{121} \\
2 o_{107} =&       - o_{36} - o_{38} + o_{45} + o_{48} \\
2 o_{108} =&   - o_{37} - o_{38} + o_{46} + o_{47} + o_{66} - o_{73} \\
2 o_{109} =&       - 3 o_{36} + o_{38} + 3 o_{45} - o_{48} \\
4 o_{111}=& - o_{36}- o_{38} +  o_{42}+  o_{44} +  o_{45} +  o_{48} -   o_{53}
  -   o_{56}-   o_{79}-   o_{81} +  o_{103}+  o_{105}-  o_{111}-  o_{113} +  o_{119}
  +  o_{121} \\
4 o_{119} =&         - 3 o_{79} - 3 o_{81} + o_{119} + o_{121} \\
4 o_{120} =&  - 3 o_{57} + o_{63} - 3 o_{80} - 3 o_{82} + o_{120} + o_{122} \\
4 o_{121} =&       - 9 o_{79} + 3 o_{81} + 3 o_{119} - o_{121} \\
4 o_{122} =&  3 o_{57} - o_{63} - 3 o_{80} - 3 o_{82} + o_{120} + o_{122} \\
16 o_{137} =& - 2 o_{13} - 2 o_{14} - 2 o_{17} - 2 o_{18} + o_{36} + o_{38} + o_{39} + o_{41}
 + 5 o_{45} + 4 o_{46} + o_{48} + 5 o_{49} \nonumber \\
&+ 4 o_{50} + o_{52} + 2 o_{95}  + 2 o_{97}
 + 4 o_{99} + 4 o_{100} + 2 o_{107} + 2 o_{109} + 4 o_{111} + 4 o_{112} \nonumber \\
&- 2 o_{137}
 - 2 o_{140} - 2 o_{141} - 2 o_{142} \\
16 o_{140} =& - 6 o_{13} + 2 o_{14} - 6 o_{17} + 2 o_{18} + 3 o_{36} - o_{38} + 3 o_{39}
 - o_{41} + 15 o_{45} - 4 o_{46} - o_{48}  \nonumber \\
&+ 15 o_{49} - 4 o_{50} - o_{52}+ 6 o_{95}
 - 2 o_{97} + 12 o_{99} - 4 o_{100} + 6 o_{107} - 2 o_{109} + 12 o_{111}\nonumber \\
& - 4 o_{112}
 - 6 o_{137} + 2 o_{140} - 6 o_{141} + 2 o_{142},
\end{align}
}
and on particles 2-3,
{\allowdisplaybreaks
\begin{align}
o_{28} =&      o_{5} - o_{6} - o_{14} + o_{15} \\
2 o_{29} =& o_{5} - o_{6} - o_{11} + o_{12} - o_{14} + o_{15} + o_{22} - o_{23} + o_{76}
                    - o_{77} + o_{96} - o_{97} + o_{100} - o_{101} \nonumber \\
&- o_{116} + o_{117} \\
2 o_{31} =&     3 o_{76} - 3 o_{77} - o_{116} + o_{117} \\
o_{60} =&       o_{37} - o_{38} - o_{46} + o_{47} \\
2 o_{61} =& o_{37} - o_{38} - o_{43} + o_{44} - o_{46} + o_{47} + o_{54} - o_{55} - o_{80}
                  + o_{82} + o_{104} - o_{106} - o_{112} \nonumber \\
&+ o_{114} + o_{120} - o_{122} \\
2 o_{63} =&       - 3 o_{80} + 3 o_{82} + o_{120} - o_{122} \\
2 o_{65} =&    - o_{2} + o_{3} - o_{22} + o_{24} \\
2 o_{66} =&  - o_{5} + o_{6} - o_{14} + o_{15} - o_{92} + o_{93} \\
2 o_{67} =&    - o_{8} + o_{9} - o_{18} + o_{19} - o_{104} + o_{105} \\
2 o_{68} =&    - o_{11} + o_{12} - o_{22} + o_{23} - o_{88} + o_{89} \\
2 o_{69} =&  - o_{5} + o_{6} - o_{14} + o_{15} + o_{92} - o_{93} \\
2 o_{70} =& - o_{8} + o_{9} - o_{18} + o_{19} + o_{104} - o_{105} \\
2 o_{71} =&   - o_{11} + o_{12} - o_{22} + o_{23} + o_{88} - o_{89} \\
2 o_{72} =& - o_{2} + o_{3} - o_{14} + o_{16} - o_{18} + o_{20} + o_{22} - o_{24} + o_{80}
                                                        - o_{81} - o_{84} + o_{85} \\
2 o_{73} =& - o_{2} + o_{3} - o_{14} + o_{16} - o_{18} + o_{20} + o_{22} - o_{24} - o_{80}
                                                        + o_{81} + o_{84} - o_{85} \\
2 o_{74} =&            - 3 o_{2} + 3 o_{3} + o_{22} - o_{24} \\
4 o_{79} =&- o_{13} - o_{15} + o_{17} + o_{19} - o_{79} - o_{82} - o_{83} - o_{86} \\
4 o_{82} =& - 3 o_{13} + o_{15} + 3 o_{17} - o_{19} - 3 o_{79} + o_{82} - 3 o_{83} + o_{86} \\
4 o_{84} =&o_{14} + o_{16} - o_{18} - o_{20} + o_{25} + o_{26} - o_{72} + o_{73} - o_{80}
                                                        - o_{81} - o_{84} - o_{85} \\
2 o_{87} =&     - o_{10} - o_{12} + o_{21} + o_{24} \\
2 o_{90} =&     - 3 o_{10} + o_{12} + 3 o_{21} - o_{24} \\
2 o_{91} =&           - o_{4} - o_{6} + o_{13} + o_{16} \\
2 o_{92} =&    - o_{5} - o_{6} + o_{14} + o_{15} - o_{66} + o_{69} \\
2 o_{94} =&    - 3 o_{4} + o_{6} + 3 o_{13} - o_{16} \\
4 o_{95} =& - o_{4} - o_{6} + o_{10} + o_{12} + o_{13} + o_{16} - o_{21} - o_{24} - o_{75}
                     - o_{78} - o_{95} - o_{98} - o_{99}\nonumber \\
& - o_{102} + o_{115} + o_{118} \\
4 o_{98} =&  - 3 o_{4} + o_{6} + 3 o_{10} - o_{12} + 3 o_{13} - o_{16} - 3 o_{21} + o_{24}
     - 3 o_{75} + o_{78} - 3 o_{95} + o_{98}\nonumber \\
& - 3 o_{99} + o_{102} + 3 o_{115} - o_{118} \\
2 o_{103} =&   - o_{7} - o_{9} + o_{17} + o_{20} \\
2 o_{104} =&  - o_{8} - o_{9} + o_{18} + o_{19} - o_{67} + o_{70} \\
2 o_{106} =&       - 3 o_{7} + o_{9} + 3 o_{17} - o_{20} \\
4 o_{115} =&   - 3 o_{75} - 3 o_{78} + o_{115} + o_{118} \\
4 o_{116} =&  3 o_{25} - o_{31} - 3 o_{76} - 3 o_{77} + o_{116} + o_{117} \\
4 o_{117} =&- 3 o_{25} + o_{31} - 3 o_{76} - 3 o_{77} + o_{116} + o_{117} \\
4 o_{118} =&       - 9 o_{75} + 3 o_{78} + 3 o_{115} - o_{118} \\
4 o_{127} =&  - o_{127} - o_{129} - o_{134} - o_{136} \\
2 o_{128} =&      - o_{128} - o_{135} \\
4 o_{129} =&   - 3 o_{127} + o_{129} - 3 o_{134} + o_{136} \\
2 o_{130} =&     - o_{130} - o_{132} \\
4 o_{131} =&    - 2 o_{131} - 2 o_{133} + o_{144} \\
2 o_{137} =&  - o_{137} - o_{139} \\
4 o_{138} =&    - 2 o_{138} - 2 o_{140} + o_{145} \\
4 o_{141} =&   - o_{127} - o_{129} + o_{134} + o_{136} - 2 o_{141} - 2 o_{143} \\
2 o_{142} =& - o_{128} + o_{135} - 2 o_{142} \\
o_{144} =&        2 o_{131} - 2 o_{133} \\
o_{145} =&  2 o_{138} - 2 o_{140} \\
o_{146} =&  - o_{131} + o_{133} + o_{138} - o_{140}. 
\end{align}
}
The above relations have been selected according to a criterium of simplicity. Indeed, quite lengthy combinations may arise: for instance, exchanging particles 1-2 in the operator $o_{93}$ gives rise to a combinations of 63 different operators of Table~\ref{tab:operators}. All other relations are linearly dependent on the selected ones. The algebraic manipulations have been authomatized using the program FORM \cite{form}.

We now list 15 independent linear relations among the 16 operators defined in Eq.~(\ref{eq:4nlagr}), out of 96 linear relations from Fierz's reorderings of 6 pairs of the four particles:
{\allowdisplaybreaks
\begin{align}
4 O^{4N}_1 &=  - O^{4N}_{1} - O^{4N}_{2} - O^{4N}_{4} - O^{4N}_{5} \\
4 O^{4N}_2 &= - 3 O^{4N}_{1} + O^{4N}_{2} - 3 O^{4N}_{4} + O^{4N}_{5} \\
4 O^{4N}_3 &=  - 3 O^{4N}_{2} + O^{4N}_{3} - 3 O^{4N}_{7} + O^{4N}_{8} \\
2 O^{4N}_4 &=  - O^{4N}_{4} - O^{4N}_{6} \\
2 O^{4N}_6 &= O^{4N}_{6} - 3 O^{4N}_{2} \\
4 O^{4N}_7 &=   - 2 O^{4N}_{6} - 2 O^{4N}_{7} - O^{4N}_{16} \\
4 O^{4N}_8 &= - 9 O^{4N}_{2} + 3 O^{4N}_{3} + 3 O^{4N}_{7} - O^{4N}_{8} \\
4 O^{4N}_9 &= - 3 O^{4N}_{2} - 3 O^{4N}_{3} + O^{4N}_{7} - O^{4N}_{8} + 2 O^{4N}_{9} \\
4 O^{4N}_{10} &=   - 3 O^{4N}_{4} - 3 O^{4N}_{7} + O^{4N}_{10} + O^{4N}_{11} \\
4 O^{4N}_{11} &= - 9 O^{4N}_{4} + 3 O^{4N}_{7} + 3 O^{4N}_{10} - O^{4N}_{11} \\
2 O^{4N}_{12} &= O^{4N}_{12} - 3 O^{4N}_{6} \\
4 O^{4N}_{13} &=  - 9 O^{4N}_{5} + 3 O^{4N}_{8} + 3 O^{4N}_{11} - O^{4N}_{13} \\
4 O^{4N}_{14} &=  - 3 O^{4N}_{5} + 3 O^{4N}_{8} - 6 O^{4N}_{9} + O^{4N}_{11} - O^{4N}_{13} + 2 O^{4N}_{14} \\
O^{4N}_{15} &=  2 O^{4N}_{5} - 2 O^{4N}_{6} \\
O^{4N}_{16} &= 2 O^{4N}_{6} - 2 O^{4N}_{7}. 
\end{align}
} 
All 16 operators in Eq.~(\ref{eq:4nlagr}) are therefore proportional to each other.


\begin{thebibliography}{100}
\bibitem{bernard95}
V.~Bernard, N.~Kaiser, U.~-G.~Meissner,
  Int.\ J.\ Mod.\ Phys.\  {\bf E4}, 193-346 (1995).
 \bibitem{vankolck99}
  U.~van Kolck,
  Prog.\ Part.\ Nucl.\ Phys.\  {\bf 43}, 337-418 (1999).
  \bibitem{bedaque02}
P.~F.~Bedaque, U.~van Kolck,
  Ann.\ Rev.\ Nucl.\ Part.\ Sci.\  {\bf 52}, 339-396 (2002).
  \bibitem{epelbaum06}
E.~Epelbaum,
  Prog.\ Part.\ Nucl.\ Phys.\  {\bf 57}, 654-741 (2006).
  \bibitem{epelbaum09}
  E.~Epelbaum, H.~-W.~Hammer, U.~-G.~Meissner,
  Rev.\ Mod.\ Phys.\  {\bf 81}, 1773-1825 (2009).
  \bibitem{weinberg90}
    S.~Weinberg,
  Phys.\ Lett.\  {\bf B251}, 288-292 (1990).
  \bibitem{weinberg91}
  S.~Weinberg,
  Nucl.\ Phys.\  {\bf B363}, 3-18 (1991).
  \bibitem{ordonez92}
   C.~Ordonez, U.~van Kolck,
  Phys.\ Lett.\  {\bf B291}, 459-464 (1992).
  \bibitem{ordonez94}
  C.~Ordonez, L.~Ray, U.~van Kolck,
  Phys.\ Rev.\ Lett.\  {\bf 72}, 1982-1985 (1994).
  \bibitem{ordonez96}
  C.~Ordonez, L.~Ray, U.~van Kolck,
  Phys.\ Rev.\  {\bf C53}, 2086-2105 (1996).
  \bibitem{vankolck94}
  U.~van Kolck,
  Phys.\ Rev.\  {\bf C49}, 2932-2941 (1994).
\bibitem{epelbaum02}
E.~Epelbaum, A.~Nogga, W.~Gloeckle, H.~Kamada, U.~G.~Meissner and H.~Witala,
  Phys.\ Rev.\  C {\bf 66}, 064001 (2002).
\bibitem{navratil}
P.~Navratil,
  Few Body Syst.\  {\bf 41}, 117 (2007).
\bibitem{gazit}
D.~Gazit, S.~Quaglioni and P.~Navratil,
  Phys.\ Rev.\ Lett.\  {\bf 103}, 102502 (2009).
\bibitem{Bedaque:1998kg}
  P.~F.~Bedaque, H.~W.~Hammer, U.~van Kolck,
  Phys.\ Rev.\ Lett.\  {\bf 82}, 463-467 (1999).
\bibitem{Bedaque:1998km}
  P.~F.~Bedaque, H.~W.~Hammer, U.~van Kolck,
  Nucl.\ Phys.\  {\bf A646}, 444-466 (1999).
\bibitem{Bedaque:1999ve}
  P.~F.~Bedaque, H.~W.~Hammer, U.~van Kolck,
  Nucl.\ Phys.\  {\bf A676}, 357-370 (2000).
\bibitem{Hammer:2000nf}
  H.~W.~Hammer, T.~Mehen,
  Nucl.\ Phys.\  {\bf A690}, 535-546 (2001).
 \bibitem{Bedaque:2002yg}
  P.~F.~Bedaque, G.~Rupak, H.~W.~Griesshammer {\it et al.},
  Nucl.\ Phys.\  {\bf A714}, 589-610 (2003).
\bibitem{Platter:2006ev}
  L.~Platter, D.~R.~Phillips,
  Few Body Syst.\  {\bf 40}, 35-55 (2006).
\bibitem{Platter:2006ad}
  L.~Platter,
  Phys.\ Rev.\  {\bf C74}, 037001 (2006).
\bibitem{3nforces}
 A.~Kievsky, M.~Viviani, L.~Girlanda and L.~E.~Marcucci,
  Phys.\ Rev.\  C {\bf 81}, 044003 (2010).
\bibitem{pvlagr}
 L.~Girlanda,
  Phys.\ Rev.\  C {\bf 77}, 067001 (2008).
\bibitem{relativity}
 L.~Girlanda, S.~Pastore, R.~Schiavilla and M.~Viviani,
  Phys.\ Rev.\  C {\bf 81}, 034005 (2010).
\bibitem{kievsky}
  A.~Kievsky,
  Phys.\ Rev.\  {\bf C60}, 034001 (1999).
\bibitem{epelbaum08}
E.~Epelbaum,
  Few Body Syst.\  {\bf 43}, 57-62 (2008).
\bibitem{bernard08}
V.~Bernard, E.~Epelbaum, H.~Krebs {\it et al.},
  Phys.\ Rev.\  {\bf C77}, 064004 (2008).
\bibitem{koike}
Y.~Koike and J.~Haidenbauer, Nucl. Phys. {\bf A463}, 365 (1987).
\bibitem{kievsky01}
A.~Kievsky, M.~Viviani, S.~Rosati,
  Phys.\ Rev.\  {\bf C64}, 024002 (2001).
\bibitem{sagara}
for a summary of the discrepancies between theory and experiment in the $3N$ systems, see for example K.~Sagara, Few Body Syst.\  {\bf 48}, 59-108 (2010).
\bibitem{glockle}
W.~Gloeckle, H.~Witala, D.~Huber {\it et al.},
  Phys.\ Rept.\  {\bf 274}, 107-285 (1996).
\bibitem{fisher}
B.~M.~Fisher, C.~R.~Brune, H.~J.~Karwowski {\it et al.},
  Phys.\ Rev.\  {\bf C74}, 034001 (2006).
\bibitem{deltuva1}
A.~Deltuva, A.~C.~Fonseca,
  Phys.\ Rev.\  {\bf C75}, 014005 (2007).
  \bibitem{deltuva2}
   A.~Deltuva, A.~C.~Fonseca,
  Phys.\ Rev.\ Lett.\  {\bf 98}, 162502 (2007).
  \bibitem{deltuva3}
    A.~Deltuva, A.~C.~Fonseca,
  Phys.\ Rev.\  {\bf C76}, 021001 (2007).
\bibitem{form} 
J.~A.~M.~Vermaseren,
  [math-ph/0010025].




\end{thebibliography}
\end{document}